\documentclass[aps,prl,showpacs,preprint,amsmath,amssymb]{revtex4-1}
\usepackage[dvipsnames,pdftex,usenames]{color} 
\usepackage[pdftex]{epsfig}
\usepackage[pdftex]{graphics}
\usepackage{epstopdf}
\usepackage{graphicx}
\usepackage{xspace}
\usepackage{bm}

\begin{document}

\preprint{}

\title{Optical pumping and non-destructive readout of a single magnetic impurity spin in an InAs/GaAs quantum dot}


\author{E. Baudin}
\author{E. Benjamin}
\author{A. Lema\^itre}
\author{O. Krebs}
\affiliation{CNRS-Laboratoire de Photonique et de Nanostructures, Route de Nozay, 91460 Marcoussis, France}


\date{\today}

\def\xp{$X^{+}$\xspace}
\def\xm{$X^{-}$\xspace}
\def\x0{$X^{0}$\xspace}
\def\a0{$A^{0}$\xspace}
\def\h{$h$\xspace}
\def\electron{$e^{-}$\xspace}
\def\sp{$\sigma^{+}$\xspace}
\def\sm{$\sigma^{-}$\xspace}
\def\da0{$|\!-\!1\rangle$\xspace}
\def\ua0{$|\!+\!1\rangle$\xspace}
\def\etal{~\textit{et al.}\xspace}
\def\ides{\textit{i.e.}\xspace}

\begin{abstract}
We report on the resonant optical pumping of the $|\!\pm\!1\rangle$ spin states of a single Mn dopant in an InAs/GaAs quantum dot embedded itself in a charge tuneable device. The experiment relies  on a ``W'' scheme of  transitions reached when a suitable longitudinal magnetic field is applied.  The optical pumping  is achieved  via the resonant excitation of the  central $\Lambda$ system at the neutral exciton \x0 energy. For a specific gate voltage,  the red-shifted photoluminescence  of the charged exciton \xm is observed, which allows non-destructive readout of the spin polarization.
 An arbitrary  spin preparation in the $|\!+\!1\rangle$ or $|\!-\!1\rangle$ state characterized by a polarization near or above 50\% is evidenced.
\end{abstract}



\maketitle

\begin{figure}
\includegraphics[width=0.8 \textwidth]{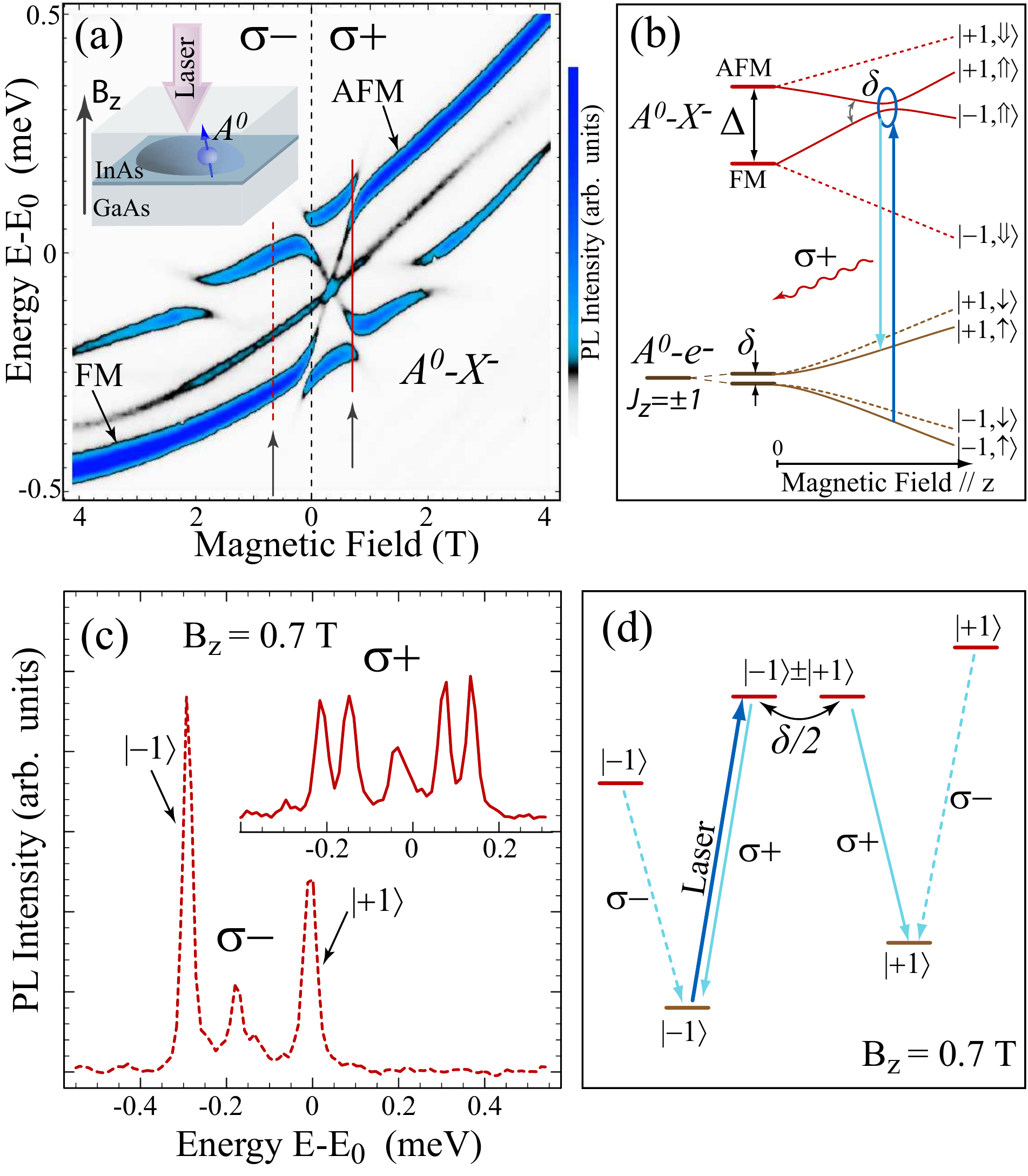}
\caption{\label{Fig1} (a) Density plot of $\mu$-PL intensity under non-resonant  excitation (HeNe at 633~nm) for a Mn-doped QD in magnetic field. Detection is \sp or \sm polarized as indicated and centered around $E_0=1.274$~eV. (b) Level diagram corresponding to the main transitions in (a) for a field below 1~T. (c) Crossed-sections of \sp and \sm spectra in (a) at  0.7~T.
(d) ``W'' transition  scheme at the field where FM and AFM levels anticross in (b). It is valid both for \a0-\x0 and \a0-\xm, hence only the selections rules (\sp or \sm) and \a0 states are indicated.}
\end{figure}

The possibility to precisely prepare and control the quantum state of an individual spin in a quantum dot (QD) is an appealing issue both from a fundamental point of view and for potential applications in the field of quantum information~\cite{Leuenberger2001}. Recent works  have demonstrated that the  spin of a an electron localized in a gate-defined or  self-assembled  semiconductor QD can be  coherently manipulated~\cite{Hanson2007,Press2008,Berezovsky2008,Kim2010}. However,  this system is not really  ideal because of its intrinsic coupling to the environment (phonons, nuclear spins or surrounding  charges) which restrains its potential for future developments. QDs doped with a single magnetic impurity like a Mn atom offer an interesting alternative in this regards~\cite{Koenraad2011}. Such isolated magnetic spin which is provided by  core electrons ($3d^5$ for Mn),  is potentially better isolated from the environment than valence or conduction electrons. Besides, thanks to the $sp$-$d$ exchange which yields  spin-dependent  sub-levels,  selective addressing by  optical methods is in principle attainable. A singly Mn-doped QD could reveal itself as particularly suitable for the realization of quantum bits.\\
\indent Among recent investigations in this direction, the  initialization of a  Mn spin  in II-VI semiconductor QDs has been demonstrated via  optical pumping (OP)~\cite{LeGall2009,Goryca2009}. In II-VI host materials,  the 5/2 Mn spin  which presents  6 distinct states is however not ideal for quantum operations. In III-V QDs the Mn effective spin  turns out to be much simpler  : a Mn atom forms indeed an acceptor  state denoted \a0 (a negative center $A_\text{Mn}^-$ and a tightly bound hole) presenting a total angular momentum reduced to $J=1$ in its  ground state~\cite{Schneider1987,Bhattacharjee1999}.  Furthermore, in a non-spherical  self-assembled QD, the state $|J_z=0\rangle$ is substantially higher in energy than the two circular states $|J_z=\pm\!1\rangle$ by a few meV~\cite{Kudelski2007,Krebs2009}. At low temperature these states $|\!\pm\!1\rangle$  define a 2-level  system, with whole occupation near unity, which we refer to as the \a0 spin in the following.  Due to the large QD-induced anisotropy~\cite{Krebs2009},  \a0 spin initialization cannot be achieved following a simple non-resonant approach  as used recently for  the ionized impurities  $A_\text{Mn}^-$ in bulk GaAs~\cite{Akimov2011,Myers2008} and based on $sp$-$d$ interaction with spin polarized electrons. An  OP scheme dedicated to the case of \a0 spin in a self-assembled QD is required.\\
\indent In this Letter, we present an OP experiment in which the \a0 spin in an InAs/GaAs QD is pumped  arbitrarily into one of its $|\!\pm\!1\rangle$ states  by using a resonant laser excitation. Its principle  relies on a ``W'' energy level scheme  which arises at a specific longitudinal magnetic  field, similarly as  quantum dot molecules in an electric field~\cite{Kim2008,Vamivakas2010}. Besides, the spin initialization and its readout are spectrally dissociated by using a technique recently  introduced by \textcite{Kloeffel2011} which enables to monitor the spin population by using a standard micro-photoluminescence ($\mu$-PL) setup.
Efficient OP is demonstrated through the vanishing (resp. enhancement) of  transitions associated to the pumped (resp.  populated) \a0 level.\\
\indent The sample used in this work  consists of a Mn-doped QD layer embedded in between an electron reservoir and a Schottky gate allowing the control of the QD charge. The  Mn doping is very dilute and less than 1\% of the QDs  are effectively doped by  a single Mn impurity, see Ref.~\cite{Kudelski2007} for details. The $\mu$-PL spectroscopy of such individual QDs is carried out  with a 2~mm focal length aspheric lens (N.A. $0.5$) actuated by piezo-stages and mounted in a split-coil magneto-optical cryostat.  The collected PL is dispersed by a 0.6 m-focal length double spectrometer and detected by a Nitrogen-cooled CCD array camera providing a $\sim 20 \mu$eV spectral resolution. All measurements presented in this Letter were performed at low temperature ($T \sim 2$ K) with a magnetic field applied parallel to the optical axis (Fig.~\ref{Fig1}(a) insert).\\
\indent The QD studied in this work contains a Mn impurity in the \a0 configuration~\cite{Schneider1987,Bhattacharjee1999,Govorov2004}. This acceptor  is characterized by an antiferromagnetic (AFM) $p$-$d$ exchange between the $3d^5$ Mn spin $S=5/2$ and the spin $J_h=3/2$ of the bound hole which gives rise to 4 levels as a function of the total angular momentum $J=1,2,3\,\text{or}\,4$. At low temperature ($T\sim2$~K) only  the ground state $J=1$ is significantly occupied so that higher states ($J=2,3,4$) can be neglected. Moreover, the lens shape and biaxial strain of InAs self-assembled QDs  shift the $|J_z=0\rangle$ state of the $J=1$ triplet  to a higher energy by about 2~meV~\cite{Krebs2009}. Consequently, this level is also negligibly occupied  at 2~K and one can  describe the   \a0 impurity as  an  effective 2-level system associated to the spin projections $J_z=\pm 1$. A natural basis to describe the \a0-QD states in presence of additional (photo-created) carriers in  the QD S-shell reads  $|\!\pm\!1,J_z^\text{QD}\rangle$, where $J_z^\text{QD}$ denotes the $z$-projection of their total spin.\\
\indent Figure~\ref{Fig1}(a) shows for the investigated QD  the typical signature of \a0 in PL spectroscopy  as a function of a positive longitudinal magnetic field.  For the gate voltage used here ($V_g=0.6$~V)  the QD is charged by an extra electron yielding  under optical excitation a negative trion \xm (a hole (\h)  with 2 electrons (\electron)). Since both conduction \electron are paired in a singlet configuration in the QD S-shell, its  spin is  determined by the  \h spin only and hence reads $J_z^\text{QD}=\pm3/2$ ($\Uparrow$ or $\Downarrow$). A detailed discussion of the spectral signature of Fig.~\ref{Fig1}(a), \ides two intense lines denoted  FM and AFM which anticross with two weaker lines forming a cross, can be found in Ref.'s \cite{Kudelski2007,Bree2008}. In brief, the   FM and AFM lines  correspond  respectively to the  ferromagnetic $|\!\pm\!1,\!\pm3/2\rangle$ and anti-ferromagnetic  $|\!\pm\!1,\!\mp3/2\rangle$ spin configurations of \a0-\xm. They are split by the exchange energy $\Delta\sim300\mu$eV, see Fig.~\ref{Fig1}(b), mostly due to the \a0-\h ferromagnetic interaction while the  \a0-\electron exchange taking place in the   final state of \xm optical recombination can be neglected~\cite{Kudelski2007}. A second very essential  feature is the anisotropic coupling $\delta/2$ within the \a0 states $|\!\pm\!1\rangle$. It gives rise to a level splitting $\delta\sim50\mu$eV at zero field both for  \a0 or \a0-\electron, and at  finite magnetic fields $B=\pm \Delta / (2 g_{A^0}\mu_\text{B})$  for \a0-\x0 or \a0-\xm states when  a FM and a AFM level  are brought in coincidence by  Zeeman effect (Fig.~\ref{Fig1}(b)).  Here, $g_{A^0}$ denotes  the effective $g$-factor of \a0 spin $J=1$ and $\mu_\text{B}$ is the Bohr magneton.
For these specific fields we identify a ``W'' transition scheme (Fig. \ref{Fig1}(d)) which, interestingly,  reproduces the same scheme as used in Ref.'s~\cite{Kim2008,Vamivakas2010} to initialize and readout non-destructively an electron spin in a QD molecule. It is composed of a central $\Lambda$ system allowing  to induce  \a0 spin flip transitions ($|\!+\!1\rangle\rightarrow|\!-\!1\rangle$ or $|\!-\!1\rangle\rightarrow|\!+\!1\rangle$) through \sp resonant excitation followed by spontaneous emission (\ides OP), and two outer \sm transitions allowing non-destructive readout of the \a0 spin as they  do not change the \a0 state $|\pm 1\rangle$ while being  cycled~\footnote{The \sm crossed transitions \mbox{$|\pm 1,\Downarrow\rangle\leftrightarrow|\mp 1,\downarrow\rangle$} are slightly permitted due to the $\delta/2$-induced coupling. However, their  oscillator strengths are
notably reduced by $\sim(\delta/2\Delta)^2$ in comparison to the direct ones.}.\\
\indent Let's first examine more closely the \a0-\xm PL spectrum under cw non-resonant and unpolarized  excitation  at the anticrossing field $B=0.7$~T (Fig.~\ref{Fig1}(c)). The \sm-resolved PL spectrum shows mainly two lines  which correspond to the W outer transitions and therefore reflect the population probability $p_{\pm1}$ of the  $|\!\pm\!1\rangle$ states. The low energy state \da0 is indeed found  more populated than the \ua0 state, yielding  a spin polarization $(p_{-1}-p_{+1})/(p_{-1}+p_{+1})$ of $\sim$20\%. However,  the associated effective temperature determined from the  PL intensity ratio  amounts to $\sim$10~K which is substantially higher than the lattice temperature~\footnote{This is also confirmed by the noticeable   central line in Fig.~\ref{Fig1}(a),(c) corresponding to the  $J_z^{A^0}=0$ level which becomes more populated than expected at 2~K.}. This effective heating is partially due to  the energy relaxation of the photo-carriers under non-resonant excitation. Yet, another key contribution comes from the   optical recombination of \sp polarized trions through the $\Lambda$ transitions which tend to equilibrate the \ua0 and \da0 populations. This mechanism   is evidenced  by the \sp-resolved PL spectrum  in Fig.~\ref{Fig1}(c) which shows two doublets of roughly equivalent intensity corresponding to both  $\Lambda$ branches. Eventually,   \sp and \sm PL spectra under non-resonant excitation are both encouraging indications to achieve OP and \a0 spin readout using this  W system.\\
\begin{figure}
\includegraphics[width=0.8 \textwidth]{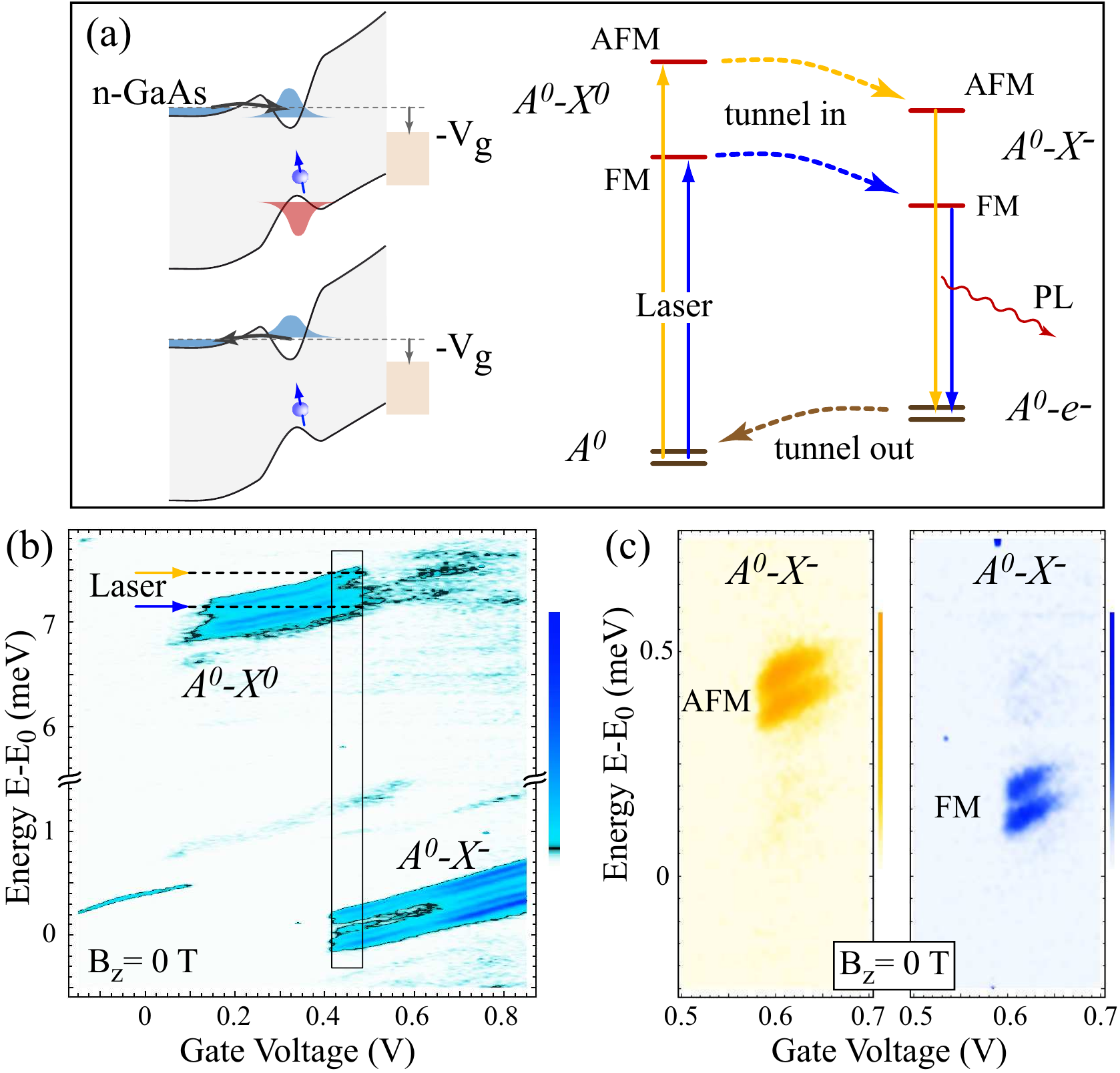}
\caption{\label{Fig2} (a) Schematics of the charge tuneable structure and of the \x0-\xm cycle used for selective resonant excitation of a specific \a0-\x0 spin configuration ($B_z=0$~T). (b) PL intensity density plot of the Mn-doped QD   against detection energy and gate voltage. The excitation   is quasi-resonant around  $E_0 + 50$~meV.  (c) Similar density plot of \xm PL observed through the resonant excitation of either the FM or AFM \x0 level in a narrow voltage range  around the \x0-\xm frontier.}
\end{figure}
\indent Obviously, non energy-selective excitation of the QD cannot lead to OP. Even with a polarized excitation  creating only \sp-polarized trions the  \da0$\rightarrow$\ua0  conversion via the $\Lambda$ transitions would be as probable as the \ua0$\rightarrow$\da0  conversion. To favor one of them, an energy selective excitation is necessary, but in that case measuring the PL signal at almost the same energy turns out somewhat problematic. To circumvent this issue we used a solution  based on a singular excitation-emission cycle that takes place in a  narrow range of applied gate voltages $V_g$ at the frontier between \x0 and \xm   \cite{Kloeffel2011,Simon2011}. In this range,  a resonantly created exciton \x0 is energetically unstable. It attracts in a time $\tau_c\sim50$~ps much shorter than the radiative lifetime an electron which tunnels in the QD from the n-GaAs reservoir. This leads to the formation of an \xm  trion which recombines optically producing a PL signal  red-shifted by $\sim$7~meV with respect to the excitation energy, after which the \electron left in the QD escapes back to the reservoir  as depicted in Fig.~\ref{Fig2}(a).  This cycle allows to resonantly excite \x0 with a laser energy outside of the spectrometer field of view and to collect the red-shifted \xm PL signal in order to probe the \a0 spin. In practice,  the  internal diffusion of the laser inside the spectrometer can still blind the detector. Using  crossed polarizers ($H/V$ or \sp/\sm) between excitation and detection  is necessary to reduce  this stray light by a factor of a few 10$^{3}$, while the remaining average background can be easily removed by taking  a reference spectrum  (with no PL signal) at a lower bias. The range of applied gate voltages for which this cycle occurs can be estimated by looking for the coexistence of \x0 and \xm spectral lines under quasi-resonant excitation (Fig.~\ref{Fig2}(b)). However, a significant voltage offset (up to +0.2~V) has to be applied when the excitation goes to resonance, to account for the reduction of  the photocarrier-induced  electric field screening.\\
\indent As shown in Fig.~\ref{Fig2}(c), by resonantly exciting the \x0  upper (lower) doublet around $V_g= 0.62$~V we are able to detect the PL signal of  the corresponding   \xm upper (lower) doublet.
The good preservation around 70\% of the  selected FM or AFM  configuration  validates our approach for  selectively   addressing and reading out \a0 spin states via the \x0-\xm cycle.
The unwished  ``cross-talking'' is  due to the large broadening of  \x0 resonance of about $100~\mu$eV\footnote{The \x0 resonance broadening, observed also for undoped QDs in this sample, results both from the \x0 lifetime shortening and from a saturating excitation power necessary to deal with a significative QD spectral wandering. We have checked this is not due to Overhauser effect as observed in Ref.~\cite{Kloeffel2011}.} rather than  a  hole or \a0 spin-flip associated to the electron tunnelling in the dot. In this regards we benefit from the fact that all exchange interactions are essentially longitudinal (Ising type) because of the heavy character of the QD and \a0 hole,  and therefore work together to conserve  a given spin orientation during the \electron capture process~\cite{Krebs2009}. Similarly, when the electron tunnels out of the dot after \xm recombination, one can reasonably expect that the
weak \a0-\electron interaction ($<30~\mu$eV) which is also mainly longitudinal, will  not imply any transitions between \ua0 and \da0  states.
 In practice, the resonant excitation of \x0  appears to be   equivalent to that of  \xm with respect to \a0 spin, as it could be anticipated from their very similar PL spectra against magnetic field~\cite{Kudelski2007}.\\
\indent Thanks to its  broad  spectral resonance the \x0-\xm  cycle   can be driven with a laser line at a fixed wavelength while sweeping the magnetic field from  0 to $\sim$1.5~T. Figures~\ref{Fig3}(b),(c) show the resulting PL density plots for a linearly polarized laser line respectively centered on the upper (AFM) or lower (FM)  doublet of \a0-\x0 in zero field. As discussed above, the detection is  linearly polarized orthogonally to that of the  excitation.  For a direct comparison,   Fig.~\ref{Fig3}(a)  displays a zoom of Fig.\ref{Fig1}(a) where  both \sp (solid lines)  and \sm (dashed lines)   measurements are reported. Around 0.7~T the W   configuration  aimed for OP  arises and  yields indeed  drastic changes of the detected PL spectra : the $\Lambda$ \sp lines corresponding to the pumped level and exhibiting the  anticrossing at 0.7~T almost completely vanish, while the outer \sm lines corresponding to the populated  \a0 spin state are  conserved or even reinforced. As sketched in the insets of Fig.~\ref{Fig3}(b),(c), the laser energy determines which transition of the $\Lambda$ is selected and in turn which \a0 spin state is pumped.  Exciting the $\Lambda$ upper or lower branch   transfers the \a0 spin  in the \ua0 state (Fig.~\ref{Fig3}(b)) or \da0 state (Fig.~\ref{Fig3}(c)). Quantitatively,  from the PL intensity ratio of the \sp and \sm lines  one can  estimate the  \a0 spin polarization achieved in both cases to  about -55\%  and  +60\% respectively.  This analysis relies   on the assumption that  the PL line intensities   correctly reflect the initial \ua0 and \da0 populations, whereas the laser is slightly blue- or red-detuned from the respective resonances. Such assumption sounds yet quite reasonable because the laser line was  centered between the \sp and \sm transitions.\\
\begin{figure}
\includegraphics[width=0.8 \textwidth]{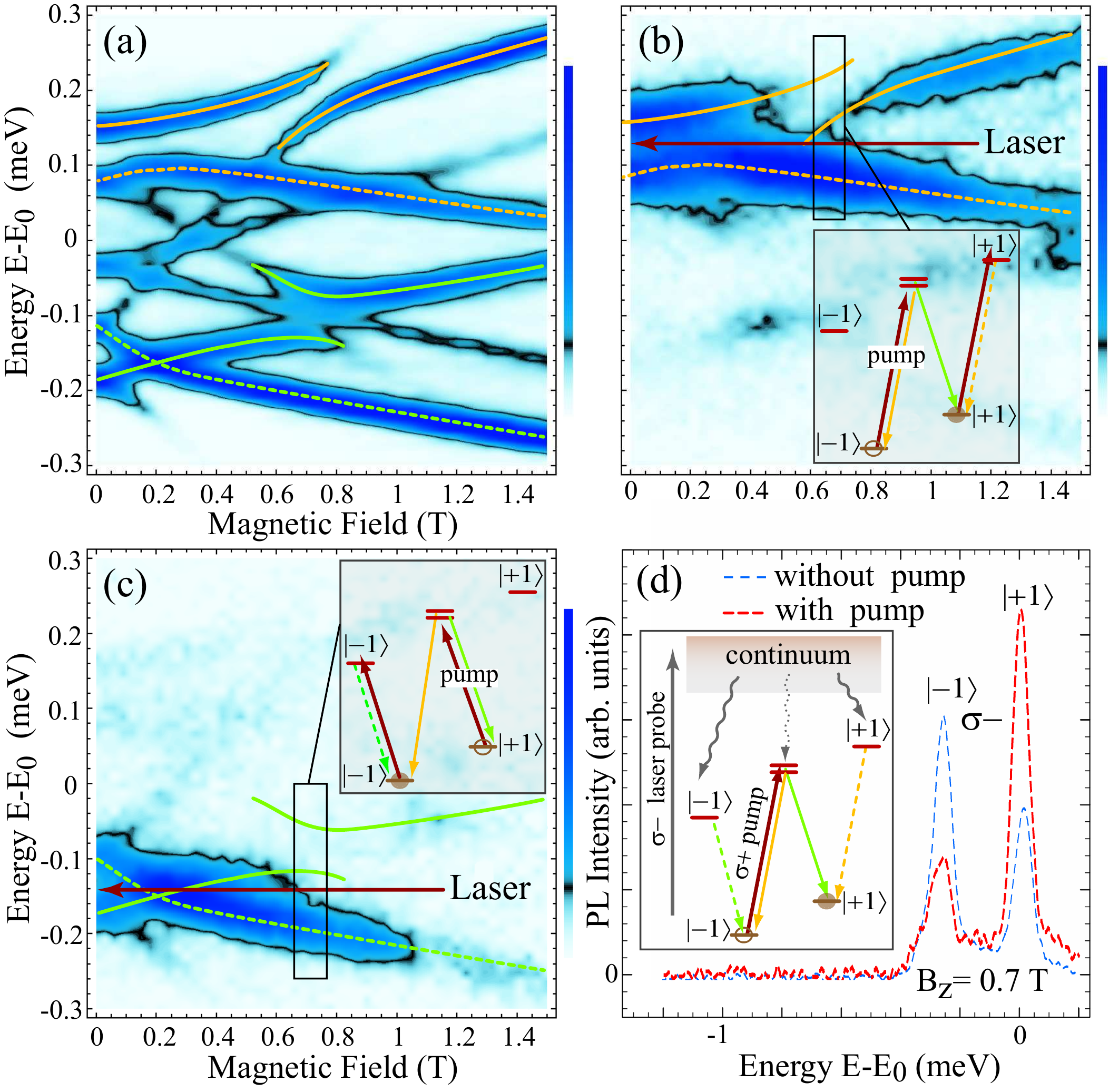}
\caption{\label{Fig3}(a) Zoom of Fig.~\ref{Fig1}(a) showing  both \sp (solid lines) and \sm (dashed lines) spectra.  (b) and (c)  PL density plots under resonant excitation centered on the zero field upper or lower  \a0-\x0 doublet. The vanishing  of the \sp anticrossing line around  0.7~T while the  \sm line persists or even increases  indicates an efficient OP mechanism as depicted in the respective insets. (d) \sm PL spectra at 0.7~T  under  \sm  quasi-resonant excitation  (at $E_0+\sim 50$~meV), with or without a  resonant \sp excitation driving the upper $\Lambda$ branch. The enhancement (reduction) of the \ua0 (\da0) line with the pump laser demonstrates an efficient OP of -49\%. }
\end{figure}
\indent To better evidence the effect of OP by using both outer branches of the W, as done to estimate the thermal polarization in Fig.~\ref{Fig1}(c), we performed a different experiment (Fig.~\ref{Fig3} (d)) : the resonant pump laser is now  \sp  circularly polarized and  drives  only the  $\Lambda$ upper branch, while a second quasi-resonant  laser (at $E_0+\sim 50$~meV) is used for the readout. Note that suppressing the pump laser stray light  requires now a \sm detection which in turn implies to use such quasi-resonant excitation for the \sm  probe laser. The main drawback of this approach is that  the \sm  polarization of the photo-created excitons or trions  is partially lost during the relaxation to the QD S-shell. This leads to some feeding of the $\Lambda$ upper state,  as illustrated in  the inset of Fig.~\ref{Fig3}(d), which tends to depolarize the \a0 spin. Nevertheless,  when turning on the pump laser, we clearly observe  OP signature  through the drastic increase (decrease) of the \ua0 (\da0) line intensity. Under these non-optimal conditions a spin polarization, clearly opposite to the thermal one, of -49\% is still reached. Conversely, by pumping the $\Lambda$ lower branch (not shown) we could increase the thermal  polarization from +21\% to  +55\%.  The unperfect excitation selectivity  associated  to the \x0-\xm cycle is likely the main limitation of the uncomplete spin preparation achieved here. Truly resonant  techniques  should be employed  to estimate properly the role of intrinsic \a0 spin relaxation.\\
\indent In this Letter we have presented the efficient optical pumping and non-destructive readout of a single magnetic spin forming a 2-level system and  embedded in an InAs/GaAs QD.
Our results are stimulating for study  further its potential as a qubit  in condensed matter. In particular, thanks to  the  QD-induced anisotropy which provides  this system with a  nano-magnet character,  one may  anticipate a long spin relaxation time.\\
\indent We acknowledge B.~Urbaszek for fruitful discussions. This  work  has  been  supported  by  French  ANR-P3N contract  QUAMOS and  the  r\'{e}gion  Ile-de-France.\\

\begin{thebibliography}{23}%
\makeatletter
\providecommand \@ifxundefined [1]{%
 \@ifx{#1\undefined}
}%
\providecommand \@ifnum [1]{%
 \ifnum #1\expandafter \@firstoftwo
 \else \expandafter \@secondoftwo
 \fi
}%
\providecommand \@ifx [1]{%
 \ifx #1\expandafter \@firstoftwo
 \else \expandafter \@secondoftwo
 \fi
}%
\providecommand \natexlab [1]{#1}%
\providecommand \enquote  [1]{``#1''}%
\providecommand \bibnamefont  [1]{#1}%
\providecommand \bibfnamefont [1]{#1}%
\providecommand \citenamefont [1]{#1}%
\providecommand \href@noop [0]{\@secondoftwo}%
\providecommand \href [0]{\begingroup \@sanitize@url \@href}%
\providecommand \@href[1]{\@@startlink{#1}\@@href}%
\providecommand \@@href[1]{\endgroup#1\@@endlink}%
\providecommand \@sanitize@url [0]{\catcode `\\12\catcode `\$12\catcode
  `\&12\catcode `\#12\catcode `\^12\catcode `\_12\catcode `\%12\relax}%
\providecommand \@@startlink[1]{}%
\providecommand \@@endlink[0]{}%
\providecommand \url  [0]{\begingroup\@sanitize@url \@url }%
\providecommand \@url [1]{\endgroup\@href {#1}{\urlprefix }}%
\providecommand \urlprefix  [0]{URL }%
\providecommand \Eprint [0]{\href }%
\providecommand \doibase [0]{http://dx.doi.org/}%
\providecommand \selectlanguage [0]{\@gobble}%
\providecommand \bibinfo  [0]{\@secondoftwo}%
\providecommand \bibfield  [0]{\@secondoftwo}%
\providecommand \translation [1]{[#1]}%
\providecommand \BibitemOpen [0]{}%
\providecommand \bibitemStop [0]{}%
\providecommand \bibitemNoStop [0]{.\EOS\space}%
\providecommand \EOS [0]{\spacefactor3000\relax}%
\providecommand \BibitemShut  [1]{\csname bibitem#1\endcsname}%
\let\auto@bib@innerbib\@empty
\bibitem [{\citenamefont {Leuenberger}\ and\ \citenamefont
  {Loss}(2001)}]{Leuenberger2001}%
  \BibitemOpen
  \bibfield  {author} {\bibinfo {author} {\bibfnamefont {M.~N.}\ \bibnamefont
  {Leuenberger}}\ and\ \bibinfo {author} {\bibfnamefont {D.}~\bibnamefont
  {Loss}},\ }\href {\doibase 10.1038/35071024} {\bibfield  {journal} {\bibinfo
  {journal} {Nature}\ }\textbf {\bibinfo {volume} {410}},\ \bibinfo {pages}
  {789} (\bibinfo {year} {2001})}\BibitemShut {NoStop}%
\bibitem [{\citenamefont {Hanson}\ \emph {et~al.}(2007)\citenamefont {Hanson},
  \citenamefont {Kouwenhoven}, \citenamefont {Petta}, \citenamefont {Tarucha},\
  and\ \citenamefont {Vandersypen}}]{Hanson2007}%
  \BibitemOpen
  \bibfield  {author} {\bibinfo {author} {\bibfnamefont {R.}~\bibnamefont
  {Hanson}}, \bibinfo {author} {\bibfnamefont {L.~P.}\ \bibnamefont
  {Kouwenhoven}}, \bibinfo {author} {\bibfnamefont {J.~R.}\ \bibnamefont
  {Petta}}, \bibinfo {author} {\bibfnamefont {S.}~\bibnamefont {Tarucha}}, \
  and\ \bibinfo {author} {\bibfnamefont {L.~M.~K.}\ \bibnamefont
  {Vandersypen}},\ }\href@noop {} {\bibfield  {journal} {\bibinfo  {journal}
  {Rev. Mod. Phys.}\ }\textbf {\bibinfo {volume} {79}},\ \bibinfo {pages}
  {1217} (\bibinfo {year} {2007})}\BibitemShut {NoStop}%
\bibitem [{\citenamefont {Press}\ \emph {et~al.}(2008)\citenamefont {Press},
  \citenamefont {Ladd}, \citenamefont {Zhang},\ and\ \citenamefont
  {Yamamoto}}]{Press2008}%
  \BibitemOpen
  \bibfield  {author} {\bibinfo {author} {\bibfnamefont {D.}~\bibnamefont
  {Press}}, \bibinfo {author} {\bibfnamefont {T.~D.}\ \bibnamefont {Ladd}},
  \bibinfo {author} {\bibfnamefont {B.}~\bibnamefont {Zhang}}, \ and\ \bibinfo
  {author} {\bibfnamefont {Y.}~\bibnamefont {Yamamoto}},\ }\href@noop {}
  {\bibfield  {journal} {\bibinfo  {journal} {Nature}\ }\textbf {\bibinfo
  {volume} {456}},\ \bibinfo {pages} {218} (\bibinfo {year}
  {2008})}\BibitemShut {NoStop}%
\bibitem [{\citenamefont {Berezovsky}\ \emph {et~al.}(2008)\citenamefont
  {Berezovsky}, \citenamefont {Mikkelsen}, \citenamefont {Stoltz},
  \citenamefont {Coldren},\ and\ \citenamefont {Awschalom}}]{Berezovsky2008}%
  \BibitemOpen
  \bibfield  {author} {\bibinfo {author} {\bibfnamefont {J.}~\bibnamefont
  {Berezovsky}}, \bibinfo {author} {\bibfnamefont {M.~H.}\ \bibnamefont
  {Mikkelsen}}, \bibinfo {author} {\bibfnamefont {N.~G.}\ \bibnamefont
  {Stoltz}}, \bibinfo {author} {\bibfnamefont {L.~A.}\ \bibnamefont {Coldren}},
  \ and\ \bibinfo {author} {\bibfnamefont {D.~D.}\ \bibnamefont {Awschalom}},\
  }\href@noop {} {\bibfield  {journal} {\bibinfo  {journal} {Science}\ }\textbf
  {\bibinfo {volume} {320}},\ \bibinfo {pages} {349} (\bibinfo {year}
  {2008})}\BibitemShut {NoStop}%
\bibitem [{\citenamefont {Kim}\ \emph {et~al.}(2010)\citenamefont {Kim},
  \citenamefont {Truex}, \citenamefont {Xu}, \citenamefont {Sun}, \citenamefont
  {Steel}, \citenamefont {Bracker}, \citenamefont {Gammon},\ and\ \citenamefont
  {Sham}}]{Kim2010}%
  \BibitemOpen
  \bibfield  {author} {\bibinfo {author} {\bibfnamefont {E.~D.}\ \bibnamefont
  {Kim}}, \bibinfo {author} {\bibfnamefont {K.}~\bibnamefont {Truex}}, \bibinfo
  {author} {\bibfnamefont {X.}~\bibnamefont {Xu}}, \bibinfo {author}
  {\bibfnamefont {B.}~\bibnamefont {Sun}}, \bibinfo {author} {\bibfnamefont
  {D.~G.}\ \bibnamefont {Steel}}, \bibinfo {author} {\bibfnamefont {A.~S.}\
  \bibnamefont {Bracker}}, \bibinfo {author} {\bibfnamefont {D.}~\bibnamefont
  {Gammon}}, \ and\ \bibinfo {author} {\bibfnamefont {L.~J.}\ \bibnamefont
  {Sham}},\ }\href@noop {} {\bibfield  {journal} {\bibinfo  {journal} {Phys.
  Rev. Lett.}\ }\textbf {\bibinfo {volume} {104}},\ \bibinfo {pages} {167401}
  (\bibinfo {year} {2010})}\BibitemShut {NoStop}%
\bibitem [{\citenamefont {Koenraad}\ and\ \citenamefont
  {Flatt\'{e}}(2011)}]{Koenraad2011}%
  \BibitemOpen
  \bibfield  {author} {\bibinfo {author} {\bibfnamefont {P.~M.}\ \bibnamefont
  {Koenraad}}\ and\ \bibinfo {author} {\bibfnamefont {M.~E.}\ \bibnamefont
  {Flatt\'{e}}},\ }\href@noop {} {\bibfield  {journal} {\bibinfo  {journal}
  {Nature Mat.}\ }\textbf {\bibinfo {volume} {10}},\ \bibinfo {pages} {91}
  (\bibinfo {year} {2011})}\BibitemShut {NoStop}%
\bibitem [{\citenamefont {{Le Gall}}\ \emph {et~al.}(2009)\citenamefont {{Le
  Gall}}, \citenamefont {Besombes}, \citenamefont {Boukari}, \citenamefont
  {Kolodka}, \citenamefont {Cibert},\ and\ \citenamefont
  {Mariette}}]{LeGall2009}%
  \BibitemOpen
  \bibfield  {author} {\bibinfo {author} {\bibfnamefont {C.}~\bibnamefont {{Le
  Gall}}}, \bibinfo {author} {\bibfnamefont {L.}~\bibnamefont {Besombes}},
  \bibinfo {author} {\bibfnamefont {H.}~\bibnamefont {Boukari}}, \bibinfo
  {author} {\bibfnamefont {R.}~\bibnamefont {Kolodka}}, \bibinfo {author}
  {\bibfnamefont {J.}~\bibnamefont {Cibert}}, \ and\ \bibinfo {author}
  {\bibfnamefont {H.}~\bibnamefont {Mariette}},\ }\href {\doibase
  10.1103/PhysRevLett.102.127402} {\bibfield  {journal} {\bibinfo  {journal}
  {Phys. Rev. Lett.}\ }\textbf {\bibinfo {volume} {102}},\ \bibinfo {pages}
  {127402} (\bibinfo {year} {2009})}\BibitemShut {NoStop}%
\bibitem [{\citenamefont {Goryca}\ \emph {et~al.}(2009)\citenamefont {Goryca},
  \citenamefont {Kazimierczuk}, \citenamefont {Nawrocki}, \citenamefont
  {Golnik}, \citenamefont {Gaj},\ and\ \citenamefont {Kossacki}}]{Goryca2009}%
  \BibitemOpen
  \bibfield  {author} {\bibinfo {author} {\bibfnamefont {M.}~\bibnamefont
  {Goryca}}, \bibinfo {author} {\bibfnamefont {T.}~\bibnamefont
  {Kazimierczuk}}, \bibinfo {author} {\bibfnamefont {M.}~\bibnamefont
  {Nawrocki}}, \bibinfo {author} {\bibfnamefont {A.}~\bibnamefont {Golnik}},
  \bibinfo {author} {\bibfnamefont {J.~A.}\ \bibnamefont {Gaj}}, \ and\
  \bibinfo {author} {\bibfnamefont {P.}~\bibnamefont {Kossacki}},\ }\href@noop
  {} {\bibfield  {journal} {\bibinfo  {journal} {Phys. Rev. Lett.}\ }\textbf
  {\bibinfo {volume} {103}},\ \bibinfo {pages} {087401} (\bibinfo {year}
  {2009})}\BibitemShut {NoStop}%
\bibitem [{\citenamefont {Schneider}\ \emph {et~al.}(1987)\citenamefont
  {Schneider}, \citenamefont {Kaufmann}, \citenamefont {Wilkening},
  \citenamefont {Baeumler},\ and\ \citenamefont {K\"{o}hl}}]{Schneider1987}%
  \BibitemOpen
  \bibfield  {author} {\bibinfo {author} {\bibfnamefont {J.}~\bibnamefont
  {Schneider}}, \bibinfo {author} {\bibfnamefont {U.}~\bibnamefont {Kaufmann}},
  \bibinfo {author} {\bibfnamefont {W.}~\bibnamefont {Wilkening}}, \bibinfo
  {author} {\bibfnamefont {M.}~\bibnamefont {Baeumler}}, \ and\ \bibinfo
  {author} {\bibfnamefont {F.}~\bibnamefont {K\"{o}hl}},\ }\href {\doibase
  10.1103/PhysRevLett.59.240} {\bibfield  {journal} {\bibinfo  {journal} {Phys.
  Rev. Lett.}\ }\textbf {\bibinfo {volume} {59}},\ \bibinfo {pages} {240}
  (\bibinfo {year} {1987})}\BibitemShut {NoStop}%
\bibitem [{\citenamefont {Bhattacharjee}(1999)}]{Bhattacharjee1999}%
  \BibitemOpen
  \bibfield  {author} {\bibinfo {author} {\bibfnamefont {A.}~\bibnamefont
  {Bhattacharjee}},\ }\href {\doibase 10.1016/S0038-1098(99)00438-X} {\bibfield
   {journal} {\bibinfo  {journal} {Sol. Stat. Comm.}\ }\textbf {\bibinfo
  {volume} {113}},\ \bibinfo {pages} {17} (\bibinfo {year} {1999})}\BibitemShut
  {NoStop}%
\bibitem [{\citenamefont {Kudelski}\ \emph {et~al.}(2007)\citenamefont
  {Kudelski}, \citenamefont {Lema\^{\i}tre}, \citenamefont {Miard},
  \citenamefont {Voisin}, \citenamefont {Graham}, \citenamefont {Warburton},\
  and\ \citenamefont {Krebs}}]{Kudelski2007}%
  \BibitemOpen
  \bibfield  {author} {\bibinfo {author} {\bibfnamefont {A.}~\bibnamefont
  {Kudelski}}, \bibinfo {author} {\bibfnamefont {A.}~\bibnamefont
  {Lema\^{\i}tre}}, \bibinfo {author} {\bibfnamefont {A.}~\bibnamefont
  {Miard}}, \bibinfo {author} {\bibfnamefont {P.}~\bibnamefont {Voisin}},
  \bibinfo {author} {\bibfnamefont {T.~C.~M.}\ \bibnamefont {Graham}}, \bibinfo
  {author} {\bibfnamefont {R.~J.}\ \bibnamefont {Warburton}}, \ and\ \bibinfo
  {author} {\bibfnamefont {O.}~\bibnamefont {Krebs}},\ }\href@noop {}
  {\bibfield  {journal} {\bibinfo  {journal} {Phys. Rev. Lett.}\ }\textbf
  {\bibinfo {volume} {99}},\ \bibinfo {pages} {247209} (\bibinfo {year}
  {2007})}\BibitemShut {NoStop}%
\bibitem [{\citenamefont {Krebs}\ \emph {et~al.}(2009)\citenamefont {Krebs},
  \citenamefont {Benjamin},\ and\ \citenamefont {Lema\^{\i}tre}}]{Krebs2009}%
  \BibitemOpen
  \bibfield  {author} {\bibinfo {author} {\bibfnamefont {O.}~\bibnamefont
  {Krebs}}, \bibinfo {author} {\bibfnamefont {E.}~\bibnamefont {Benjamin}}, \
  and\ \bibinfo {author} {\bibfnamefont {A.}~\bibnamefont {Lema\^{\i}tre}},\
  }\href@noop {} {\bibfield  {journal} {\bibinfo  {journal} {Phys. Rev. B}\
  }\textbf {\bibinfo {volume} {80}},\ \bibinfo {pages} {165315} (\bibinfo
  {year} {2009})}\BibitemShut {NoStop}%
\bibitem [{\citenamefont {Akimov}\ \emph {et~al.}(2011)\citenamefont {Akimov},
  \citenamefont {Dzhioev}, \citenamefont {Korenev}, \citenamefont {Kusrayev},
  \citenamefont {Sapega}, \citenamefont {Yakovlev},\ and\ \citenamefont
  {Bayer}}]{Akimov2011}%
  \BibitemOpen
  \bibfield  {author} {\bibinfo {author} {\bibfnamefont {I.}~\bibnamefont
  {Akimov}}, \bibinfo {author} {\bibfnamefont {R.}~\bibnamefont {Dzhioev}},
  \bibinfo {author} {\bibfnamefont {V.}~\bibnamefont {Korenev}}, \bibinfo
  {author} {\bibfnamefont {Y.}~\bibnamefont {Kusrayev}}, \bibinfo {author}
  {\bibfnamefont {V.}~\bibnamefont {Sapega}}, \bibinfo {author} {\bibfnamefont
  {D.}~\bibnamefont {Yakovlev}}, \ and\ \bibinfo {author} {\bibfnamefont
  {M.}~\bibnamefont {Bayer}},\ }\href {\doibase 10.1103/PhysRevLett.106.147402}
  {\bibfield  {journal} {\bibinfo  {journal} {Phys. Rev. Lett.}\ }\textbf
  {\bibinfo {volume} {106}},\ \bibinfo {pages} {2} (\bibinfo {year}
  {2011})}\BibitemShut {NoStop}%
\bibitem [{\citenamefont {Myers}\ \emph {et~al.}(2008)\citenamefont {Myers},
  \citenamefont {Mikkelsen}, \citenamefont {Tang}, \citenamefont {Gossard},
  \citenamefont {Flatt\'{e}},\ and\ \citenamefont {Awschalom}}]{Myers2008}%
  \BibitemOpen
  \bibfield  {author} {\bibinfo {author} {\bibfnamefont {R.~C.}\ \bibnamefont
  {Myers}}, \bibinfo {author} {\bibfnamefont {M.~H.}\ \bibnamefont
  {Mikkelsen}}, \bibinfo {author} {\bibfnamefont {J.-M.}\ \bibnamefont {Tang}},
  \bibinfo {author} {\bibfnamefont {A.~C.}\ \bibnamefont {Gossard}}, \bibinfo
  {author} {\bibfnamefont {M.~E.}\ \bibnamefont {Flatt\'{e}}}, \ and\ \bibinfo
  {author} {\bibfnamefont {D.~D.}\ \bibnamefont {Awschalom}},\ }\href@noop {}
  {\bibfield  {journal} {\bibinfo  {journal} {Nature Mat.}\ }\textbf {\bibinfo
  {volume} {7}},\ \bibinfo {pages} {203} (\bibinfo {year} {2008})}\BibitemShut
  {NoStop}%
\bibitem [{\citenamefont {Kim}\ \emph {et~al.}(2008)\citenamefont {Kim},
  \citenamefont {Economou}, \citenamefont {B\u{a}descu}, \citenamefont
  {Scheibner}, \citenamefont {Bracker}, \citenamefont {Bashkansky},
  \citenamefont {Reinecke},\ and\ \citenamefont {Gammon}}]{Kim2008}%
  \BibitemOpen
  \bibfield  {author} {\bibinfo {author} {\bibfnamefont {D.}~\bibnamefont
  {Kim}}, \bibinfo {author} {\bibfnamefont {S.}~\bibnamefont {Economou}},
  \bibinfo {author} {\bibfnamefont {S.}~\bibnamefont {B\u{a}descu}}, \bibinfo
  {author} {\bibfnamefont {M.}~\bibnamefont {Scheibner}}, \bibinfo {author}
  {\bibfnamefont {A.}~\bibnamefont {Bracker}}, \bibinfo {author} {\bibfnamefont
  {M.}~\bibnamefont {Bashkansky}}, \bibinfo {author} {\bibfnamefont
  {T.}~\bibnamefont {Reinecke}}, \ and\ \bibinfo {author} {\bibfnamefont
  {D.}~\bibnamefont {Gammon}},\ }\href {\doibase
  10.1103/PhysRevLett.101.236804} {\bibfield  {journal} {\bibinfo  {journal}
  {Phys. Rev. Lett.}\ }\textbf {\bibinfo {volume} {101}},\ \bibinfo {pages}
  {236804} (\bibinfo {year} {2008})}\BibitemShut {NoStop}%
\bibitem [{\citenamefont {Vamivakas}\ \emph {et~al.}(2010)\citenamefont
  {Vamivakas}, \citenamefont {Lu}, \citenamefont {Matthiesen}, \citenamefont
  {Zhao}, \citenamefont {F\"{a}lt}, \citenamefont {Badolato},\ and\
  \citenamefont {Atat\"{u}re}}]{Vamivakas2010}%
  \BibitemOpen
  \bibfield  {author} {\bibinfo {author} {\bibfnamefont {A.~N.}\ \bibnamefont
  {Vamivakas}}, \bibinfo {author} {\bibfnamefont {C.-Y.}\ \bibnamefont {Lu}},
  \bibinfo {author} {\bibfnamefont {C.}~\bibnamefont {Matthiesen}}, \bibinfo
  {author} {\bibfnamefont {Y.}~\bibnamefont {Zhao}}, \bibinfo {author}
  {\bibfnamefont {S.}~\bibnamefont {F\"{a}lt}}, \bibinfo {author}
  {\bibfnamefont {A.}~\bibnamefont {Badolato}}, \ and\ \bibinfo {author}
  {\bibfnamefont {M.}~\bibnamefont {Atat\"{u}re}},\ }\href@noop {} {\bibfield
  {journal} {\bibinfo  {journal} {Nature}\ }\textbf {\bibinfo {volume} {467}},\
  \bibinfo {pages} {297} (\bibinfo {year} {2010})}\BibitemShut {NoStop}%
\bibitem [{\citenamefont {Kloeffel}\ \emph {et~al.}(2011)\citenamefont
  {Kloeffel}, \citenamefont {Dalgarno}, \citenamefont {Urbaszek}, \citenamefont
  {Gerardot}, \citenamefont {Brunner}, \citenamefont {Petroff}, \citenamefont
  {Loss},\ and\ \citenamefont {Warburton}}]{Kloeffel2011}%
  \BibitemOpen
  \bibfield  {author} {\bibinfo {author} {\bibfnamefont {C.}~\bibnamefont
  {Kloeffel}}, \bibinfo {author} {\bibfnamefont {P.}~\bibnamefont {Dalgarno}},
  \bibinfo {author} {\bibfnamefont {B.}~\bibnamefont {Urbaszek}}, \bibinfo
  {author} {\bibfnamefont {B.}~\bibnamefont {Gerardot}}, \bibinfo {author}
  {\bibfnamefont {D.}~\bibnamefont {Brunner}}, \bibinfo {author} {\bibfnamefont
  {P.}~\bibnamefont {Petroff}}, \bibinfo {author} {\bibfnamefont
  {D.}~\bibnamefont {Loss}}, \ and\ \bibinfo {author} {\bibfnamefont
  {R.}~\bibnamefont {Warburton}},\ }\href {\doibase
  10.1103/PhysRevLett.106.046802} {\bibfield  {journal} {\bibinfo  {journal}
  {Phys. Rev. Lett.}\ }\textbf {\bibinfo {volume} {106}},\ \bibinfo {pages}
  {046802} (\bibinfo {year} {2011})}\BibitemShut {NoStop}%
\bibitem [{\citenamefont {Govorov}(2004)}]{Govorov2004}%
  \BibitemOpen
  \bibfield  {author} {\bibinfo {author} {\bibfnamefont {A.~O.}\ \bibnamefont
  {Govorov}},\ }\href@noop {} {\bibfield  {journal} {\bibinfo  {journal} {Phys.
  Rev. B}\ }\textbf {\bibinfo {volume} {70}},\ \bibinfo {pages} {35321}
  (\bibinfo {year} {2004})}\BibitemShut {NoStop}%
\bibitem [{\citenamefont {van Bree}\ \emph {et~al.}(2008)\citenamefont {van
  Bree}, \citenamefont {Koenraad},\ and\ \citenamefont
  {Fern\'{a}ndez-Rossier}}]{Bree2008}%
  \BibitemOpen
  \bibfield  {author} {\bibinfo {author} {\bibfnamefont {J.}~\bibnamefont {van
  Bree}}, \bibinfo {author} {\bibfnamefont {P.~M.}\ \bibnamefont {Koenraad}}, \
  and\ \bibinfo {author} {\bibfnamefont {J.}~\bibnamefont
  {Fern\'{a}ndez-Rossier}},\ }\href@noop {} {\bibfield  {journal} {\bibinfo
  {journal} {Phys. Rev. B}\ }\textbf {\bibinfo {volume} {78}},\ \bibinfo
  {pages} {165414} (\bibinfo {year} {2008})}\BibitemShut {NoStop}%
\bibitem [{Note1()}]{Note1}%
  \BibitemOpen
  \bibinfo {note} {The $\sigma ^{-}$\protect \xspace crossed transitions
  \unhbox \voidb@x \hbox {$|\pm 1,\delimiter "322B37F \delimiter "526930B
  \leftrightarrow |\mp 1,\delimiter "3223379 \delimiter "526930B $} are
  slightly permitted due to the $\delta /2$-induced coupling. However, their
  oscillator strengths are notably reduced by $\sim (\delta /2\Delta )^2$ in
  comparison to the direct ones.}\BibitemShut {Stop}%
\bibitem [{Note2()}]{Note2}%
  \BibitemOpen
  \bibinfo {note} {This is also confirmed by the noticeable central line in
  Fig.~\ref {Fig1}(a),(c) corresponding to the $J_z^{A^0}=0$ level which
  becomes more populated than expected at 2~K.}\BibitemShut {Stop}%
\bibitem [{\citenamefont {Simon}\ \emph {et~al.}(2011)\citenamefont {Simon},
  \citenamefont {Belhadj}, \citenamefont {Chatel}, \citenamefont {Amand},
  \citenamefont {Renucci}, \citenamefont {Lemaitre}, \citenamefont {Krebs},
  \citenamefont {Dalgarno}, \citenamefont {Warburton}, \citenamefont {Marie},\
  and\ \citenamefont {Urbaszek}}]{Simon2011}%
  \BibitemOpen
  \bibfield  {author} {\bibinfo {author} {\bibfnamefont {C.-M.}\ \bibnamefont
  {Simon}}, \bibinfo {author} {\bibfnamefont {T.}~\bibnamefont {Belhadj}},
  \bibinfo {author} {\bibfnamefont {B.}~\bibnamefont {Chatel}}, \bibinfo
  {author} {\bibfnamefont {T.}~\bibnamefont {Amand}}, \bibinfo {author}
  {\bibfnamefont {P.}~\bibnamefont {Renucci}}, \bibinfo {author} {\bibfnamefont
  {A.}~\bibnamefont {Lemaitre}}, \bibinfo {author} {\bibfnamefont
  {O.}~\bibnamefont {Krebs}}, \bibinfo {author} {\bibfnamefont {P.~A.}\
  \bibnamefont {Dalgarno}}, \bibinfo {author} {\bibfnamefont {R.~J.}\
  \bibnamefont {Warburton}}, \bibinfo {author} {\bibfnamefont {X.}~\bibnamefont
  {Marie}}, \ and\ \bibinfo {author} {\bibfnamefont {B.}~\bibnamefont
  {Urbaszek}},\ }\href@noop {} {\bibfield  {journal} {\bibinfo  {journal}
  {Phys. Rev. Lett.}\ }\textbf {\bibinfo {volume} {106}},\ \bibinfo {pages}
  {166801} (\bibinfo {year} {2011})}\BibitemShut {NoStop}%
\bibitem [{Note3()}]{Note3}%
  \BibitemOpen
  \bibinfo {note} {The $X^{0}$\protect \xspace resonance broadening, observed
  also for undoped QDs in this sample, results both from the $X^{0}$\protect
  \xspace lifetime shortening and from a saturating excitation power necessary
  to deal with a significative QD spectral wandering. We have checked this is
  not due to Overhauser effect as observed in Ref.~\cite
  {Kloeffel2011}.}\BibitemShut {Stop}%
\end{thebibliography}
%

\end{document}